\documentclass[
reprint,
superscriptaddress,
showpacs,
preprintnumbers,
nofootinbib,
nobibnotes,
amsmath,
amssymb, 
aps,
prd,
floatfix
]{revtex4-1}

\usepackage[utf8]{inputenc}
\usepackage[normalem]{ulem}
\usepackage{graphicx}
\usepackage{dcolumn}
\usepackage[colorlinks=true,allcolors=purple]{hyperref}
\usepackage{url}
\usepackage{enumerate}

\usepackage{slashed,multirow,relsize,soul,feynmp-auto,tikz}
\usepackage{color}
\usepackage{mathrsfs} 
\usepackage{amsmath}
\usepackage{cancel}
\usepackage{bbold}
 \usepackage{mathrsfs}
\usepackage{braket}
\usepackage{physics}
\usepackage{multirow}
\usepackage[capitalize]{cleveref}

\usepackage{fontawesome} 
\definecolor{blue-violet}{rgb}{0.33, 0.17, 0.89}

\newcounter{CommentCount}
\definecolor{MH}{rgb}{0.0,0.6,9}
\definecolor{palatinate}{rgb}{0.494, 0.192, 0.482}

\begin{document}

\title{Pion decay constraints on exotic 17 MeV vector bosons}

\author{Matheus Hostert}
\affiliation{School of Physics and Astronomy, University of Minnesota, Minneapolis, MN 55455, USA}
\affiliation{William I. Fine Theoretical Physics Institute, School of Physics and Astronomy, University of
Minnesota, Minneapolis, MN 55455, USA}
\affiliation{Perimeter Institute for Theoretical Physics, Waterloo, ON N2J 2W9, Canada}

\author{Maxim Pospelov}
\affiliation{School of Physics and Astronomy, University of Minnesota, Minneapolis, MN 55455, USA}
\affiliation{William I. Fine Theoretical Physics Institute, School of Physics and Astronomy, University of
Minnesota, Minneapolis, MN 55455, USA}

\date{\today}

\begin{abstract}
We derive constraints on the couplings of light vector particles to all first-generation Standard Model fermions using leptonic decays of the charged pion, $\pi^+\to e^+ \nu_e X_\mu$.
In models where the net charge to which $X_\mu$ couples to is not conserved, no lepton helicity flip is required for the decay to happen, enhancing the decay rate by factors of $\mathcal{O}(m_\pi^4/m_e^2m_X^2)$.
A past search at the SINDRUM-I spectrometer severely constrains this possibility.
In the context of the hypothesized $17$~MeV particle proposed to explain anomalous $^8$Be, $^4$He, and $^{12}$C nuclear transitions claimed by the ATOMKI experiment, this limit rules out vector-boson explanations and poses strong limits on axial-vector ones.
\end{abstract}

\maketitle

\section{Introduction} 

The discovery of a new vector particle would provide a strong indication that the Standard Model (SM) gauge group is incomplete.
The most minimal interpretation for such a new particle would be that of a mediator of a new local U$(1)_{Q_X}$ gauge symmetry, associated with some set of SM or dark sector quantum numbers $Q_X$.
Numerous examples of such theories have been studied in the literature, with particular emphasis on the cases of a secluded gauge boson ({\em i.e.} an additional $U(1)$ coupled to the hypercharge)~\cite{Holdom:1985ag,Fayet:1990wx,Pospelov:2008zw,Fabbrichesi:2020wbt} and the gauging of conserved vector currents, such as $Q_{\rm SM} = Q_{B-L}$
, where $B$ and $L$ stand for baryon and lepton numbers, respectively\footnote{Strictly speaking, $B-L$ is only conserved if neutrinos are Dirac particles.}.
While these options present a minimal path to a renormalizable and gauge invariant theory, one can still entertain the possibility of gauging a current that is not conserved as long as the model is seen as a low-energy effective field theory (EFT)~\cite{Adler:1969gk,Bell:1969ts}.
The scale $\Lambda$ of the EFT is then set by the mass of the new particles that restore gauge invariance.
This is the case in theories where one gauges $Q_{\rm SM} \in \{Q_{B}, Q_{L}, Q_{L_e - L_\mu, \dots}\}$, where, for instance, the gauging of $Q_B$ and $Q_L$ is anomalous, while $Q_{L_e - L_\mu}$ is known to be violated at the classical level by small effects from neutrino masses and mixing.
In these scenarios, the mass of the vector boson is necessarily non-zero, with a lower bound set by the couplings and scale of the EFT~\cite{Preskill:1990fr}.
The longitudinal mode of the mediator, then, participates in processes of energies below $\Lambda$, enhancing it by factors of $\mathcal{O}(E^2/m_X^2)$, where $E$ is the typical energy of the process~\cite{Dror:2017ehi,Dror:2017nsg}. 
Therefore, when considering gauge extensions of the SM, a departure from current conservation is usually accompanied by strong experimental limits from longitudinal mode emission. 
In the case of anomalous currents, such as those coupled to $Q_L$ or $Q_B$, the problematic amplitudes will be induced at a loop level. On the other hand, currents that are broken by tree-level effects will manifest similar enhancement already at tree level \cite{Karshenboim:2014tka,Dror:2017nsg}. 

In the MeV scale, the new vector boson, $X_\mu$, can be produced in rare or otherwise-forbidden particle decays, mediate long-range interactions, and contribute to precision observables like the $(g-2)$ of leptons.
In particular, the emission of $X_\mu$ in nuclear de-excitations has recently gained significant attention due to an apparent excess of $e^+e^-$ pairs in several nuclear transitions in the ATOMKI experiment.
The collaboration first reported the observation of a $>5\sigma$-statistically-significant excess of $e^+e^-$ pairs with large opening angles in magnetic transitions of $^8$Be$(17.6)$ and $^8$Be$(18.15)$~\cite{Krasznahorkay:2015iga,Krasznahorkay:2018snd}.
The results are inconsistent with electromagnetic (EM) internal pair creation from virtual photons, which predicts a smooth and rapidly falling distribution of opening angles.
They concluded that the feature could be explained by the decays of a slowly-moving vector boson of mass $m_X \sim 17$~MeV, mimicking the signal via the decay chain $^8$Be$^* \to ^8$Be$ \, X \to ^8$Be$ \, e^+e^-$.
Since the excess was first observed in a $J^\pi = 1^+ \to 0^+$ nuclear transition, the new boson was compatible with either a pseudoscalar ($0^-$), vector ($1^-$), or axial-vector ($1^+$) particle.
Subsequent excesses in the de-excitation of two overlapping $^4$He resonances ($0^{+(-)} \to 0^+$)~\cite{Krasznahorkay:2019lyl,Krasznahorkay:2021joi} and in $^{12}$C ($1^-\to 0^+$)~\cite{Krasznahorkay:2022pxs} indicate that the spin and parity of the hypothetical $X(17)$ boson must be either of a vector or axial-vector nature~\cite{Feng:2020mbt}. 
While this article is about new-physics explanations for these anomalous decays, we note that the most pressing issue currently is whether the ATOMKI results can be independently reproduced in other experiments.

The candidate models for $17$~MeV vector or axial-vector boson have been discussed at length in the literature~\cite{Feng:2016jff,Kozaczuk:2016nma,Feng:2016ysn}.
Note that the possibility of a pseudoscalar is incompatible with the anomaly claimed in $^{12}$C transition~\cite{Krasznahorkay:2022pxs}.
This includes the theoretically-motivated case of a MeV-scale QCD axion~\cite{Alves:2017avw,Alves:2020xhf}, which we previously pointed out as an ideal target for multi-lepton searches in meson decays~\cite{Hostert:2020xku}\footnote{A recent NA62 search for $K^+\to \pi^+ a(17) a(17)$ has, in fact, ruled out the QCD axion interpretation~\cite{na62search}.}.
The vector explanation also predicted a proton-energy-independent excess coming from continuum emission of $X_\mu$~\cite{Zhang:2020ukq}.
While this excess was not present in the original measurements, the collaboration later reported it after a revised background evaluation~\cite{Sas:2022pgm}.
In addition, a vector particle is only viable if the $^4$He excess is exclusively from the $0^+ \to 0^+$ transition.
While the vector couplings needed to explain the Be and He transitions are compatible~\cite{Feng:2020mbt}, they are in a $\gtrsim 4\sigma$ tension with those preferred by the excess in carbon.
In the case of axial-vectors, the nuclear uncertainties are much more significant, and the tension has not been quantified~\cite{Barducci:2022lqd}.
Contrary to the vector case, the uncertainty in the axial-vector transition matrix elements does not cancel in the relevant ratios, such as $(^8{\rm Be}^* \to ^8{\rm Be} X)/(^8{\rm Be}^* \to ^8{\rm Be} \gamma)$.
Similarly to the pseudoscalar scenario, an axial-vector particle has also been invoked to explain the discrepancy in $\pi^0 \to e^+e^-$~\cite{Kahn:2007ru}, where the theoretical predictions~\cite{Dorokhov:2007bd,Vasko:2011pi,Husek:2014tna,Hoferichter:2021lct,Christ:2022rho} are in a $2 - 3\sigma$ tension with the KTEV measurement~\cite{KTeV:2006pwx}. 

\begin{figure}[t]
    \centering
\includegraphics[width=0.45\textwidth]{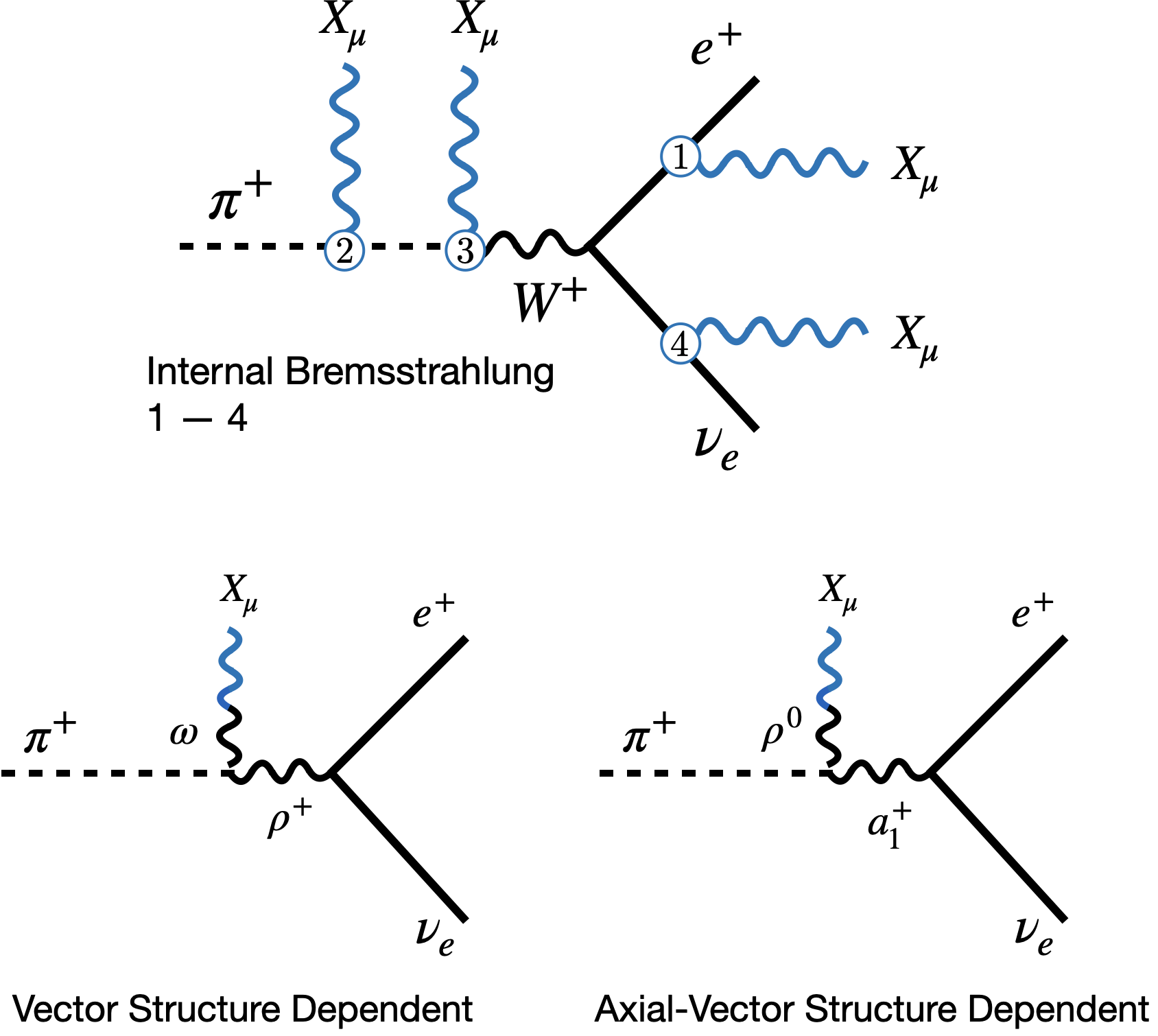}
    \caption{The four internal $X$ bremsstrahlung amplitudes (left) and the vector and axial structure dependent amplitudes (right). 
    The mixing mediates the latter two between $X$ and the vector mesons and is subdominant.
    \label{fig:diagrams}}
\end{figure}

In this article, we confront the existence of $X(17)$ with leptonic decays of the charged pion.
In the SM, all the dominant decay modes of the charged pion are suppressed by the lepton mass due to the pseudoscalar nature of the pion, and, in the case of radiative decays, due to gauge invariance (for reviews, see~\cite{Bryman:1982et} and \cite{Piilonen:1986bv}). 
Specifically, in the limit $m_\pi/m_\rho \to 0$, the amplitude for {\em any} electron decay mode, $\pi^+\to e^+\nu_e,\,e^+\nu_e\gamma,\,e^+\nu_e e^+e^-$, will be suppressed by small value of $m_e$. 
However, in beyond-the-SM scenarios, the emission of $X$ can be enhanced if it is allowed to couple to non-conserved currents. 
This lifts the lepton mass suppression and enhances the rate, potentially by a factor as large as $m_\pi^4/(m_e^2m_X^2)$,
allowing to set strong limits on models of light particles~\cite{Dror:2017nsg,Dror:2020fbh,Hostert:2020xku,Altmannshofer:2022izm}.

\begin{figure}[t]
    \centering
    \includegraphics[width=0.49\textwidth]{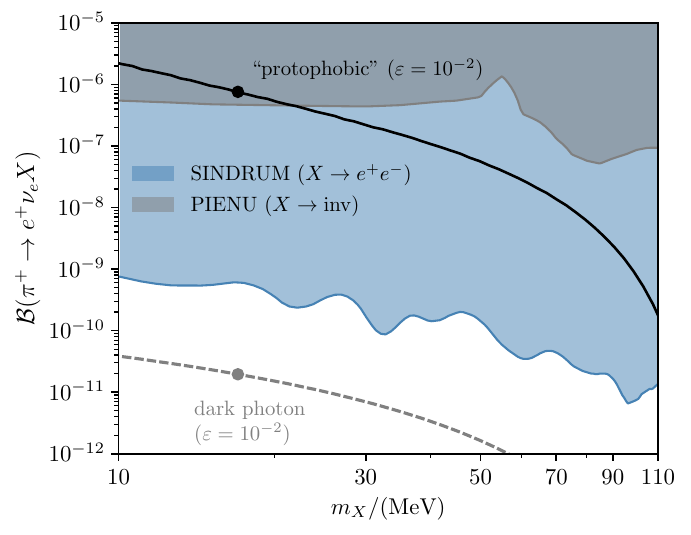}
    \caption{The 90\% C.L. limits on $\pi^+\to e^+ \nu_e X$ decays from SINDRUM, valid for $\mathcal{B}(X\to e^+e^-) = 1$, and PIENU, valid for $\mathcal{B}(X\to {\rm inv}) = 1$.
    Also shown are the total branching ratio predictions for a protophobic vector boson ($Q_p^V = Q_\nu^V = 0$ and $Q_e^V \ll Q_n^V = 1$) and a dark photon, both for $\varepsilon = 10^{-2}$.
    \label{fig:br_limits}}
\end{figure}

Our main result is that vector ($1^-$) explanations of the ATOMKI results are excluded by the existing experimental limits on leptonic pion decays, $\pi^+ \to e^+ \nu_e X$. 
The diagrams responsible for such a decay are shown in \cref{fig:diagrams}. 
Some axial-vector explanations ($1^+$) remain compatible~\cite{Kozaczuk:2016nma,Barducci:2022lqd} but are strongly constrained.
In particular, we show that the vector boson model proposed to evade limits from $\pi^0 \to X \gamma$ and $e^+e^- \to X \gamma$, usually referred to as a \emph{protophobic} vector boson, is robustly excluded. 
The $\pi^+$ decay branching ratio predicted for the best-fit point of the $^8$Be anomaly explanation \cite{Feng:2016jff} is excluded by over three orders of magnitude, as shown in \cref{fig:br_limits}.

\section{Vector and Axial-Vector X(17)}
\label{sec:models}

We focus on a light vector boson coupled to SM fermions.
We start from a generic St\"ueckelberg theory for $X_\mu$,
\begin{equation}
    \mathscr{L} \supset -\frac{1}{4} X_{\mu \nu}X^{\mu \nu} + \frac{m_X^2}{2} X_\mu X^\mu + e \varepsilon X_\mu \mathcal{J}^\mu_X,
\end{equation}
where the current involves only SM fermions and has the form,
\begin{equation}\label{eq:current}
    \mathcal{J}^\mu_X =   \sum_{f = \{e, u, d, \nu\}} \overline{f} \gamma^\mu \left( Q_f^V + Q_f^A \gamma^5 \right) f,
\end{equation}
with $Q_f^V$ and $Q_f^A$ the vector and axial-vector charges of the fermions $f$ under the new force. We will treat these interactions in the low-energy limit, where the theoretical consistency of these models is not fully apparent. 
The requirements for the corresponding currents to be conserved impose very strong constraints on admissible terms in (\ref{eq:current}).
The conservation of $\mathcal{J}^\mu_X$ will constrain the charge assignment to be vectorial and, in particular, a linear combination of the conserved currents in the SM, $\mathcal{J}_{\rm EM}$ and $\mathcal{J}_{B-L}$.
These correspond to the two well-known cases of a kinetically-mixed dark photon and the mediator of a local $U(1)_{\rm B-L}$ gauge symmetry.
The purely baryonic current, $\mathcal{J}_{\rm B}$, is broken by the chiral anomaly at the loop level and will require the entire ``anomalon" sector above the electroweak scale to make these anomalies cancel (see {\em e.g.} \cite{Dobrescu:2014fca}).
Apart from such flavor-universal assignments, leptonic assignments like $L_\alpha - L_\beta$ are also conserved, up to small effects from neutrino masses (see, e.g., Ref.~\cite{Dror:2020fbh}).
If all $Q^V_f$ correspond to the EM charges of the SM particles, and all $Q^A_f=0$, then $\varepsilon$ can be identified with the so-called kinetic mixing parameter of a dark photon model. Axial-vector currents are explicitly not conserved in the presence of fermion masses, which implies serious problems for a model if couplings to heavier fermions are on the same order of magnitude as to lighter ones. Since our main motivation is the ongoing experimental anomaly, our approach in this paper is to ignore these higher-level problems of theoretical consistency and address low-energy phenomenology (pion/nuclear physics) using (\ref{eq:current}) as input. 
In the same vein, we allow either vector or axial-vector coupling to electrons and quarks to avoid an excessive amount of parity violation mediated by light particles \cite{Bouchiat:2004sp}, noting that $Q^V_e Q^A_{u(d)}$ combination is less constrained than $Q^A_eQ^V_{u(d)}$ and $Q^A_eQ^V_{e}$.

The couplings in \cref{eq:current} can be used to find the coupling of $X$ to nucleons, $N=\{p,n\}$, and to the charged pion, $\pi^+$. 
In the case of vector couplings, they are given by the charges $Q_p^V = 2Q_u^V + Q_d^V$, $Q_n^V = Q_u^V + 2 Q_d^V$, and $Q_\pi^V = Q_u^V - Q_d^V = Q_p^V - Q_n^V$.
In the case of the axial-vector, the couplings to nucleons can be obtained using the individual up- and down-quark axial-vector matrix elements.
The couplings $\kappa_N \overline{N} \gamma^\mu \gamma^5 N$ is then given by $\kappa_p = g_A^u Q_u^A + g_A^d Q_d^A$ and $\kappa_n = g_A^d Q_u^A + g_A^u Q_d^A$.
From the neutron $\beta$-decay and the latest Lattice QCD results~\cite{Alexandrou:2019brg,Alexandrou:2021wzv}, $g_A^u = 0.817$ and $g_A^d = - 0.450$.
In the case of the pion, there is no spin for the axial-vector to couple to, but as we show in \cref{app:x17_chpt}, there is still a coupling between the pion, the $W^\pm$ boson, and the axial-vector $X_\mu$.

While most experimental limits on light vector bosons are usually presented on the parameter space of a kinetically-mixed dark photon or $U(1)_{B-L}$ gauge boson, they can easily reinterpret them as constraints on the charge assignment in \cref{eq:current}.
As recognized in Ref.~\cite{Feng:2016jff}, the combination of constraints below requires $X(17)$ to be \emph{protophobic}, $-6.7\% < Q_p/Q_n < 7.8\%$, and implies the coupling to neutrons dominates the anomalous nuclear transitions.
We briefly review existing limits on the new couplings below. 

\paragraph{Limits on $Q_{p,n}^V$:}
NA48 looked for new visible resonances in neutral pion decays~\cite{NA482:2015wmo} , $\pi^0 \to \gamma (X \to e^+e^-)$, constraining the anomaly factor $(q_u Q_u^V - q_d Q_d^V)$, requiring $\left|Q_p^V \varepsilon \right| < 8 \times 10^{-4}$~\cite{Feng:2016jff} (several other limits exist, albeit for $m_X \gtrsim 17$~MeV~\cite{SINDRUMI:1992xmn,WASA-at-COSY:2013zom,HADES:2013nab,PHENIX:2014duq}).
Since no axial-axial-vector anomalies exist, axial-vector bosons do not contribute to $\pi^0 \to X \gamma$ decays at an appreciable level.
The vector coupling to neutrons is independently constrained by the limits on new long-range interactions between neutrons and large nuclei~\cite{Barbieri:1975xy}, $|Q_n \varepsilon| < 0.025$~\cite{Feng:2016jff}.

\paragraph{Limits on $Q_e^{A, V}$:}
KLOE-2 constrained the production of $X$ associated with initial-state-radiation in $e^+e^-$ collisions, $e^+e^- \to \gamma (X\to e^+e^-)$, providing an upper limit of $\sqrt{(Q_e^V)^2 + (Q_e^A)^2} \varepsilon \lesssim 2\times 10^{-3}/\sqrt{\mathcal{B}_{ee}}$~\cite{Anastasi:2015qla}.
Lower limits can be obtained from direct searches for the decays-in-flight of $X(17)$.
The NA64 fixed-target experiment constrains $|Q_e \varepsilon| > 6.8\times10^{-4}$ ~\cite{NA64:2018lsq,NA64:2019auh} and the E141 beam-dump experiment~\cite{Riordan:1987aw,Andreas:2012mt} result gives $|Q_e \varepsilon| > 2\times10^{-4}$, where both limits assume $\mathcal{B} (X \to e^+e^-) = 1$.
In both cases, the new particle would be produced by bremsstrahlung in the interactions of an electron beam with target nuclei, $e^- Z \to (X\to e^+e^-) e^- Z$.
Because these searches take place in higher-energy beams ($E_e \sim 100$~GeV for NA64 and $E_e \sim 9$~GeV for E141), these limits are typically more stringent than the lifetime requirements of the ATOMKI anomaly, where $X$ is produced with velocity $\beta_X \sim 0.1 - 0.6$.

\paragraph{Limits on $Q_\nu^{V,A}$:}
Precision measurements of the elastic scattering of reactor antineutrino on electrons at the TEXONO experiment~\cite{TEXONO:2009knm}, $\overline{\nu}_e -e^-$, constrain $ -3 \times10^{-4} < Q_\nu Q_e^V \varepsilon^2 < 7\times 10^{-5}$~\cite{Denton:2023gat}.
In addition, the observation of coherent elastic neutrino-nucleus scattering (CEvNS) at reactors constrains $|Q_\nu Q_n^V \varepsilon| < 1.4 \times 10^{-5}$~\cite{Denton:2023gat}.
For a discussion of the constraints from neutrino experiments, see Ref.~\cite{Denton:2023gat}.
In summary, the stronger-than-Weak interactions required to explain the ATOMKI results cannot be present in the neutrino sector.
Therefore, the branching ratio $\mathcal{B}_{ee} \equiv \mathcal{B}(X \to e^+e^-)$ will be unity as long as a $Q_\nu = 0$, provided $X_\mu$ has no additional dark sector decay modes.

Note that protophobic vectors are incompatible with a $B-L$ charge assignment ($Q_p^V = Q_n^V = -Q_e^V = -Q_\nu$) as well as with a kinetically mixed dark photon ($Q_p^V = - Q_e^V$ and $Q_n^V = Q_\nu =0$).
Linear combinations of conserved currents, like $\mathcal{J}_{B-L} - \kappa \mathcal{J}_{\rm EM}$, where $\kappa$ controls the mixing between the two currents, also do not work as in that case $Q_p^V = - Q_e^V = (1-\kappa)$ and $Q_n^V = - Q_\nu  = 1$, in conflict with the constraints from the neutrino sector.
In addition, it is also easy to see that no combination of conserved currents can accommodate the ATOMKI result without violating at least one of the electron, neutrino, or $\pi^0$ limits.
As already identified by the literature on the topic~\cite{Feng:2016jff,Kozaczuk:2016nma,Barducci:2022lqd}, we must consider non-conserved currents instead, which, as we show next (\cref{sec:piondecays}), are strongly constrained by helicity-unsuppressed charged pion decays.

\section{Charged pion decays}
\label{sec:piondecays}

\begin{figure}[t]
    \centering
    \includegraphics[width=0.49\columnwidth]{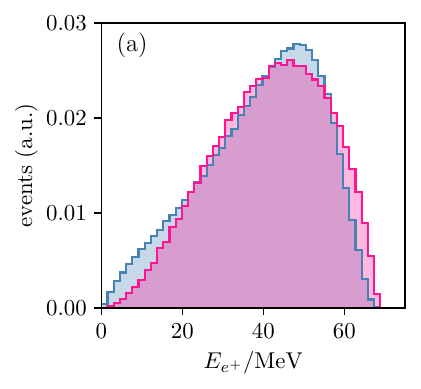}
    \includegraphics[width=0.49\columnwidth]{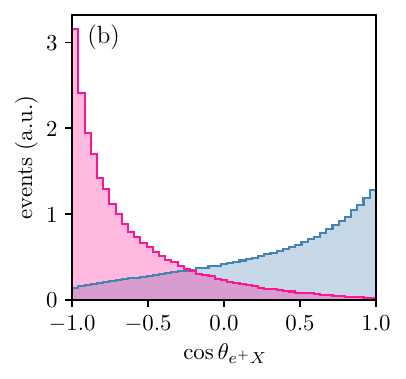}
    \includegraphics[width=0.49\columnwidth]{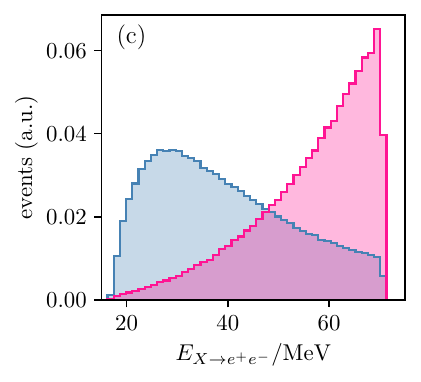}
    \includegraphics[width=0.49\columnwidth]{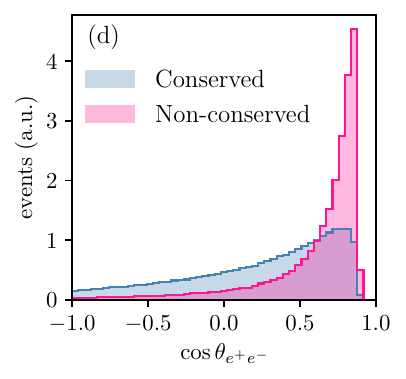}
    \caption{The normalized event distributions of $\pi^+ \to e^+ \nu_e X$ in the pion rest frame for both conserved (dark photon) and non-conserved currents (e.g., protophobic vector). We show a) the energy of the primary positron, b) the cosine of its angle with $X$, c) the total $X$ energy, and d) the cosine of the opening angle of the secondary $e^+e^-$.
    \label{fig:distributions}}
\end{figure}

In this section, we discuss the rate of $X$ emission in leptonic meson decays, $\pi^+(k_1) \to e^+(k_2) \, \nu_e (k_3) \, X(k_4)$.
The pion interaction Lagrangian is given by
\begin{align}\label{eq:pionLagrangian} 
\mathscr{L} &\supset  i e \varepsilon X^\mu \left(  Q_3^V \pi^+ \overset{\leftrightarrow}{\partial_\mu} \pi^- + g V_{ud} F_\pi Q^R_3W_{\mu}^+ \pi^-\right) + \text{h.c.},
\end{align}
where $Q_3^i = (Q_u^i - Q_d^i)/2$ is the isovector quark charge under the $U(1)_X$, $F_\pi$ is the pion decay constant, $V_{ud}$ the CKM matrix element, and $g$ the weak coupling. The right-handed coupling is defined as $Q^R_i = Q^V_i+Q^A_i$.
The second term is the contact interaction between $X_\mu$, the pion, and the $W^\pm_\mu$ bosons.
For the SM photon, this term is guaranteed by gauge invariance and is responsible for the exact cancellation of helicity-unsuppressed amplitudes in radiative pion decay.
\Cref{eq:pionLagrangian,eq:current} uniquely determine the internal bremsstrahlung (IB) contribution to the decays of the point-like pion.
The interactions of $X_\mu$ with pions in Chiral Perturbation theory are given in \cref{app:x17_chpt} and the full pion decay IB amplitude is given in \cref{app:amplitude}.

Away from the point-like limit, photons can be emitted from strongly interacting states inside the pion through the structure-dependent (SD) emission.
Under the vector-meson dominance framework, this contribution takes place by the emission of a virtual pair of vector mesons, such as $\rho$ and $\omega$ or $\rho$ and $a_1$.
In this case, a vector meson converts to the photon and the other vector or axial-vector meson then mediates the leptonic interaction.
As such, this contribution is not suppressed by a helicity flip but by the masses of the vector mesons.
The SD component in QED can only be observed at large $e^+-\gamma$ angles, where the IB amplitude is small.
For the $X$ boson, the SD diagrams will depend on the couplings of $X$ to quarks and its mixing with the lightest vector and axial-vector mesons (see~\cref{app:amplitude}).
Since SD terms do not dominate the rate even for conserved currents or in QED, they can be safely neglected in our discussion.

The contribution of vertex (\ref{eq:pionLagrangian}) should be combined with the emission of $X$ from the external legs of the underlying $\pi^+ \to e^+\nu_e$ decay.
By explicit calculation, the helicity-unsuppressed amplitude for pion decay to a $X_\mu$ boson of polarization $s$ is
\begin{equation} 
\mathcal{M}_\mu (\varepsilon^\mu_s)^* = \Delta Q_X \, \sqrt{2} V_{ud}^* G_F f_\pi \,\overline{u}(k_3) \gamma_\mu P_L v(k_2) (\varepsilon^\mu_s)^*.
\end{equation}
As expected, when the $Q_X$ charge is conserved, $\Delta Q_X \equiv Q_u^R - Q_d^R - Q_\nu^L + Q_e^L = 0$, this amplitude vanishes.
Away from that limit, the emission of transverse and longitudinal modes of $X_\mu$ takes place.
Note that helicity-suppression is necessarily present for $X_\mu$ coupled exclusively to right-handed leptonic currents.
A similar conclusion was reached for axion-like particles in Ref.~\cite{Altmannshofer:2022izm}.
The amplitude is still helicity-suppressed at the classical level for $B$ and $L$ gauge bosons. 
Neither of these two options, however, are suitable for explaining the ATOMKI anomalies. 

To shed more light on the underlying enhancement of the decay rate, we decompose the differential pion decay rate in transverse and longitudinal mode emission. 
In the rest frame of the pion, the angles of the two leptons with respect to the vector boson, $\theta_{e X}$ and $\theta_{\nu X}$, are given by
\begin{align}
 \lambda_e &\equiv \sin^2(\theta_{e X}/2) = (1-z)/xy,
 \\
 \lambda_\nu &\equiv \sin^2 (\theta_{\nu X}/2) = (1-y)/xz,
\end{align}
where $x = 2 (k_1 \cdot k_4)/m_\pi^2 = 2E_X/m_\pi$, $y = 2 (k_1 \cdot k_2)/m_\pi^2 = 2 E_e / m_\pi$, and $z = 2 (k_1 \cdot k_3)/m_\pi^2 = 2 E_\nu/ m_\pi = 2 - x -y$ are the usual kinematic variables.
We neglected higher order terms in $m_e/m_\pi$ and $m_X/m_\pi$.

The differential rates for the emission of left-handed (LH), right-handed (RH), and longitudinal (L) $X_\mu$ bosons are then given by,
\begin{align}\label{eq:diff_rates}
    \frac{1}{\Gamma_0} \frac{\dd \Gamma_{\rm IB}^{\rm T}}{\dd x \, \dd y} &= 
    \alpha \varepsilon^2 (\Delta Q_X)^2 \, y\,z\, \times 
    \begin{cases}
        \lambda_\nu (1-\lambda_e), & \text{if RH}\\
        \lambda_e (1-\lambda_\nu), & \text{if LH}\\
    \end{cases}
    \\ 
    \frac{1}{\Gamma_0} \frac{\dd \Gamma_{\rm IB}^{\rm L}}{\dd x \, \dd y} &= 
    \alpha \varepsilon^2 (\Delta Q_X)^2 \,\frac{m_\pi^2}{m_X^2} x^2  y\,z  \frac{\lambda_\nu\lambda_e}{2},
\end{align}
where $\Gamma_0 \equiv \frac{G_F^2 m_\pi^3 F_\pi^2}{16\pi^2}$. 
Longitudinal modes are preferentially emitted in the opposite direction of both leptons,
while transverse modes are preferentially emitted along the direction of one of the leptons.
Helicity suppression is absent in all cases, showing that the non-conservation of $Q_X$ charge has significant consequences for longitudinal \emph{as well as} transverse mode emission.
The $E_X^2/m_X^2$ enhancement factor is a consequence of the absence of a Ward identity, $\mathcal{M}_\mu k_4^\mu \neq 0$, signaling the breakdown of gauge invariance.

\begin{figure}[t]
    \centering
    \includegraphics[width=0.49\textwidth]{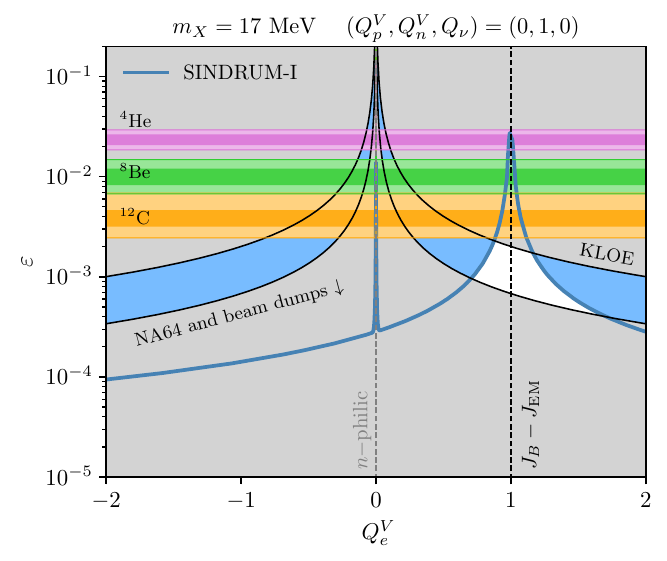}
    \caption{Constraints on the $X$ coupling to SM fermions as a function of the electron charge, $Q_e^V$, for the limit of a protophobic vector boson, $Q_p^V = 2 Q_u^V + Q_d^V = 0$.
    In blue, we show the region excluded by the SINDRUM-I measurement of $e^+e^-$ pairs in leptonic pion decay.
    The colorful horizontal bands show the $1\sigma$ and $2\sigma$ preference regions for the ATOMKI results.    \label{fig:Qe_limits}}
\end{figure}

One advantage of considering pion decays is that they are independent of the couplings of $X_\mu$ to the second or third generation, providing a more direct probe of the ATOMKI results.
Theoretically and experimentally, the internal bremsstrahlung part of the amplitude is well-understood, so small deviations from the SM can be easily identified, especially when searching for a visible resonance.
In principle, $K^+$, $D^+$, and $B^+$ mesons may also provide useful constraints~\cite{Reece:2009un,Ilten:2015hya,Chiang:2016cyf,Ibe:2016dir,Krnjaic:2019rsv,Castro:2021gdf}, although in those cases, the experimental activity is largely focused on the structure-dependent rates, and searches for new resonances are not available at such low invariant masses.
In what follows, we only discuss the experimental limits on exotic pion decays.

\subsubsection{SINDRUM-I limits}

SINDRUM-I was a spectrometer with $4\pi$ coverage originally designed to search for $\mu^+ \to e^+ e^+e^-$ at the Paul Scherrer Institute (PSI)~\cite{Eichler:2021efn}.
Because of its tracking capabilities, it was also able to perform the most precise measurements of $\pi^+ \to e^+ \nu_e e^+e^-$ and $\mu^+ \to \overline{\nu}_\mu \nu_e e^+ e^+e^-$ to date and provides the best limits on exotic three-track decays of charged pions, $\pi^+\to e^+ \nu_e X$, with $X\to e^+e^-$.
With a total of $4\times10^{12}$ pion decays, SINDRUM-I constrained the branching ratio to light scalar particles in the interval of $10$~MeV $<m_X< 110$~MeV to be below $\mathcal{O}(10^{-9})$ and $\mathcal{O}(10^{-11})$~\cite{SINDRUM:1989qan}.
The signal was simulated using the differential decay rate to a light Higgs particle, which has similar kinematics to the emission of the longitudinal mode of $X_\mu$.
We have checked that this difference in the kinematics is not substantial and that a light Higgs would display very similar properties to the pink histograms shown in \cref{fig:distributions}.

The resulting limit for a $17$~MeV boson is $\mathcal{B}(\pi^+ \to e^+ \nu_e X_{ee} | m_X = 17$ MeV$) < 6.0 \times 10^{-10}$ at 90\% C.L.
Considering only the helicity-unsuppressed longitudinal emission of $X_\mu$, the SINDRUM-I constraints can be translated into a limit on $Q_X$ charge conservation as
\begin{equation}
    e\varepsilon |\Delta Q_X|  < \frac{8.5 \times 10^{-5}}{\sqrt{B_{\rm ee}}} \text{  at  } 90\% \text{ C.L.}
\end{equation}

The analysis also required a time coincidence between the electron and positron, requiring $(\delta t)^2 = \left[(t_1^+ - t_-)^2 + (t_1^+ - t_2^+)^2\right]/2 < (600$~ps)$^2$.
In principle, since the electron is produced from the decay of $X_\mu$, it may be delayed with respect to the primary positron.
We consider this when drawing our limits, although it does not impose a significant constraint on the parameter space of interest.
For the protophobic vector with $Q_\nu = 0$, the typical lifetime of $X$ is
\begin{equation}
    \tau_X 
    \simeq 0.2 \text{ fs } \left(\frac{(Q_e^V \varepsilon)^2 + (Q_e^A\varepsilon)^2}{10^{-4}}\right)^{-1},
\end{equation}
much below the experimental timing resolution.

\begin{figure*}[t]
    \centering
    \includegraphics[width=0.49\textwidth]{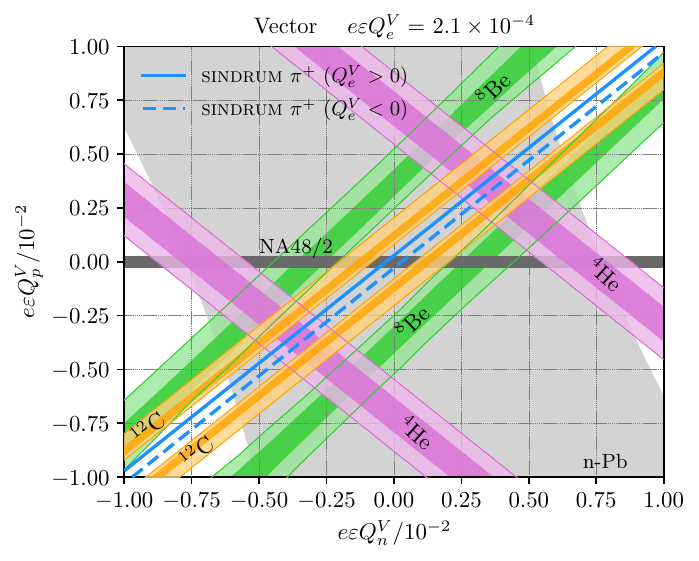}
    \includegraphics[width=0.49\textwidth]{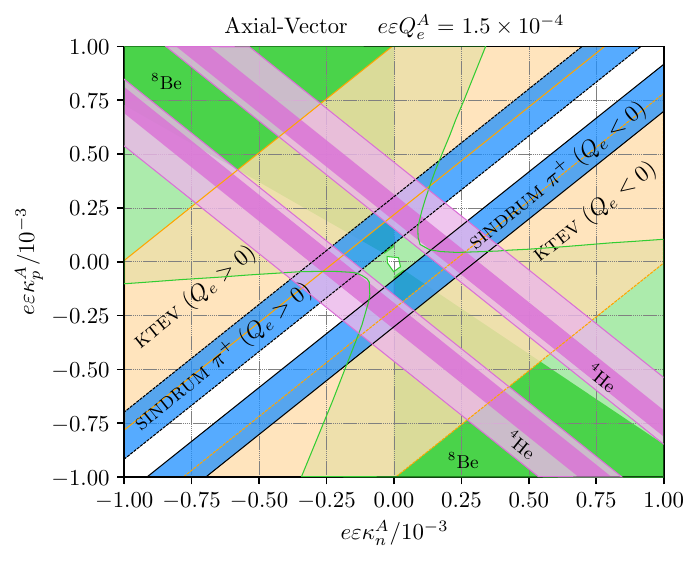}
    \caption{The $X_\mu$ couplings to nucleons for a fixed coupling to electrons for a vector (left) and axial-vector (right) boson.
    Filled regions indicate allowed parameter space. The region allowed by the SINDRUM-I search for $\pi^+ \to e^+ \nu (X \to e^+e^-)$ is shown in blue and is thinner than the line width for the vector case.
    ATOMKI regions of preference are shown at the $1\sigma$ ($2\sigma$) level as dark (light) colorful bands.
    \label{fig:final_limits}}
\end{figure*}

\subsubsection{PIENU limits}

If $X_\mu$ possesses an invisible branching ratio, a complementary limit can be derived from a search for $\pi^+ \to e^+ \nu X_{\rm inv}$ at PIENU~\cite{PIENU:2021clt}.
The PIENU detector studied pion decays at TRIUMF, using a calorimeter to measure the positron energy from pion and secondary muon decays.
Based on the energy spectrum of the positrons observed from a total of $1.3 \times 10^{6}$ $\pi^+ \to e^+ \nu_e$ events, the search in Ref.~\cite{PIENU:2021clt} set a limit on the branching ratio of the three-body decay mode of $\mathcal{B}(\pi^+ \to e^+ \nu_e X_{\rm inv} | m_X = 17$ MeV$) < 4.7 \times 10^{-7}$ at 90\% C.L.
The kinematics of the decay, in particular the positron energy distribution, is again very similar to the signal considered by the collaboration.
In terms of the couplings of the $17$~MeV boson and its invisible branching ratio, the PIENU limit is
\begin{equation}
    e\varepsilon |\Delta Q_X|  < \frac{2.5 \times 10^{-3}}{\sqrt{B_{\rm inv}}}  \text{  at  } 90\% \text{ C.L.}
\end{equation}
As expected, this limit is much weaker than the visible one.
In addition, if the invisible branching ratio is dominated by the decay into neutrinos, much stronger limits from neutrino scattering will apply.

\section{Discussion}

This section discusses the impact of the charged pion decay limits on the vector and axial-vector interpretations of the ATOMKI results.
A phenomenological fit to the ATOMKI results in these models was recently performed in Ref.~\cite{Barducci:2022lqd}, and we will base our discussion around their results.
We use the latest fits from~\cite{BarducciErratum}.
For the $^8$Be decays, we neglect isospin-breaking effects ($\xi = 0$ in Ref.~\cite{Barducci:2022lqd}), although the resulting exclusion of the preferred regions remains strong when including them.
The typical values required to explain the anomaly are $\varepsilon \sim \mathcal{O}(10^{-3} - 10^{-2})$ for the vector case and $\varepsilon \sim \mathcal{O}(10^{-4} - 10^{-3})$ for the axial-vector one.
The latter are subject to more significant uncertainties, and the lack of data on the $^{12}$C transition matrix element prevented the authors of \cite{Barducci:2022lqd} from drawing a region of preference for this case.
The uncertainty in the nuclear transition elements is significant for axial-vectors and does not cancel in the ratios to the measured electromagnetic matrix elements.
Simplified fits in the parameter space of $X(17)$ exist in the literature, but to our knowledge, a detailed nuclear physics study has only been performed for $^4$He~\cite{Viviani:2021stx}.
The latter study agrees with the vector boson results in \cite{Feng:2016jff,Feng:2016ysn,Feng:2020mbt}.

First, we discuss the protophobic vector proposed in~\cite{Feng:2016jff}.
Working in the limit where $2Q_u^V = -Q_d^V$ ($Q_p^V = 0$) and $Q_\nu = 0$, we show the current constraints and regions of preference for the ATOMKI results in \cref{fig:Qe_limits}, as a function of the electron coupling $Q_e^V$.
The ATOMKI results are ruled out by SINDRUM-I, except in the region where $\Delta Q_X = - Q_n^V + Q_e^V \to 0$.
Where that happens, however, the constraints on the electron coupling coming from KLOE-2 cover the entire $^8$Be and $^4$He regions of preference.

We now discuss the vector and axial-vector models away from the protophobic limit. 
The allowed regions in the parameter space of $X_\mu$ are shown in \cref{fig:final_limits} following the fit in Ref.~\cite{Barducci:2022lqd}.
In the vector case, we fix the coupling to electrons to the smallest allowed value from the NA64 limits.
The region allowed by pion decay constraints is smaller than the thickness of our lines and strongly contradicts the ATOMKI results.
Choosing the largest value of $Q_e^V$ compatible with the KLOE-2 constraints does not qualitatively change this.
No overlap with the $^{12}$C region of preference is seen for any allowed value of $Q^V_e$.
In the axial-vector case, we fix the electron coupling according to \cite{Barducci:2022lqd}, motivated by a measured electron $(g-2)$ discrepancy.
The latter also fixes the region preferred by the $\pi^0 \to e^+e^-$ anomaly, denoted by KTEV in the figure.
While we do not attempt to precisely quantify the tension between all the different results in \cref{fig:final_limits}, we note that the in the vector model, it is beyond the $4\sigma$ level with the simplified fits of Ref.~\cite{Barducci:2022lqd}.
This is driven by two main factors: the incompatibility between the $^{12}$C results and the $^{8}$Be and $^{4}$He anomalies, and the incompatibility of all the ATOMKI anomalies with both neutral and charged pion decays.
A more precise estimate of the internal tension of the model requires a more detailed description of the nuclear physics involved and more information on the experimental systematic uncertainties at ATOMKI, both currently lacking.

At this point, we can conclude that the following observations cannot be simultaneously satisfied:
\begin{enumerate}
    \item evidence for $X(17)$ in $^8$Be, $^4$He, and $^{12}$C nuclear transitions~\cite{Krasznahorkay:2015iga,Krasznahorkay:2018snd,Krasznahorkay:2019lyl,Krasznahorkay:2021joi,Krasznahorkay:2022pxs}, \label{item1}
    \item limits on $\pi^+ \to e^+ \nu_e X_{ee}$ at SINDRUM-I~\cite{SINDRUM:1989qan}, \label{item2}
    \item limits on $\pi^+ \to e^+ \nu_e X_{\rm inv}$ at PIENU~\cite{PIENU:2021clt}, \label{item3}
    \item limits on $\pi^0 \to \gamma X_{ee}$ at NA48~\cite{NA482:2015wmo}, \label{item4}
    \item limits on $e^+e^- \to \gamma X_{ee}$ at KLOE-2~\cite{Anastasi:2015qla}, \label{item5}
    \item limits on $e^- Z \to X_{ee} e^- Z$ at NA64 and beam dumps~\cite{NA64:2018lsq,NA64:2019auh,Riordan:1987aw,Andreas:2012mt}. \label{item6}
\end{enumerate}
If one further imposes the constraint that the theory is renormalizable and gauge invariant ($X$ coupled to a conserved current such as $\mathcal{J}_{B-L}$, $\mathcal{J}_{\rm EM}$, and linear combinations thereof), then the internal tension is even more severe as at least two of the aforementioned limits will be in direct contradiction with the $X(17)$ hypothesis. 
We are then forced to relax the theoretical constraint instead to include effective theories, where $X$ couples to non-conserved currents at low energies. 
In that case, the \emph{protophobic} vector, with $2Q_u^V = -Q_d^V$, has been more successful at evading experimental limits~\cite{Feng:2016jff}.
Nevertheless, given the hierarchy of charges required by \cref{item4,item5,item6},
the model does not escape pion decay constraints, specifically \cref{item2,item3} (cf. \cref{fig:br_limits}).
This adds to the internal tension in the model stemming from the three separate results in \cref{item1}, namely the different couplings preferred by the Be, He, and C results.
If $X(17)$ possesses an invisible branching ratio, the internal tension becomes even more severe, as the required couplings to explain the ATOMKI results are larger and constraints from neutrino-electron and neutrino-nucleus scattering become prohibitively strong.
Pion decays, therefore, provide an independent and robust exclusion of the protophobic scenario.

While this work does not fully exclude the small couplings behind an axial-vector explanation of ATOMKI, it poses the strongest constraints in the parameter space. 
The allowed regions are constrained to four islands in \cref{fig:final_limits}.
We emphasize that nuclear uncertainties are the most significant in this case and that no ATOMKI fit has been performed.
In addition, axial-vector bosons with the couplings allowed by our constraints can only be seen as an effective theory as the axial-vector current is not conserved.
Such explanations would necessarily have to involve quark flavor non-universality. 
Otherwise, $O(10^{-4})$ coupling to top quarks would result in a very large loop-induced $X_\mu \bar s \gamma^\mu b $ interaction and be strongly constrained by, {\em e.g.}, $B\to K^*X_\mu$ decays \cite{Dror:2017nsg}. 
It is unclear whether such models would have any reasonable UV completion that restores current conservation.

\section{Conclusions}

We find that vector- and axial-vector-boson explanations for the excess of $e^+e^-$ events at the ATOMKI experiment are significantly constrained by radiative pion decays. 
By explicitly calculating the emission of $X_\mu$ in charged pion decays, we showed that searches for a visible resonance in three-track pion decays at SINDRUM-I exclude the vector-boson interpretation of ATOMKI and significantly constrain the parameter space allowed in an axial-vector one.
In the vector case, this limit adds to the internal tension between the initial $^8$Be and $^4$He results and the more recent claim of an anomaly in $^{12}$C.
For axial-vectors, with the current fits found in the literature and keeping in mind the more significant uncertainties, we find that there are still regions of parameter space compatible with pion decays.
Given that scalar particles cannot mediate the ATOMKI transitions and that pseudoscalar particles are not compatible with the latest results in $^{12}$C, we are led to conclude that, as far as new-physics explanations of the anomaly go, these axial-vector solutions are the only possibility that is still allowed by data.
If this explanation stands, it must also be interpreted as a low-energy effective theory, and additional constraints from flavor-changing neutral currents would need to be evaluated.

If the evidence for $X(17)$ persists in the data and shows up also in other experimental setups, such as at the Montreal X17~\cite{Azuelos:2022nbu} and new JEDI~\cite{Bastin:2023utm} projects, it would be worthwhile to reconsider a $\pi^+\to e^+ \nu_e X_{ee}$ search in modern experimental setups.
To that end, a search at the PIONEER experiment at PSI could be performed, albeit with limited tracking capabilities~\cite{PIONEER:2022yag}.
An alternative would be to consider kaon factories as a secondary source of pions. 
The modern photon vetoes and tracking capabilities of NA62 could help reject backgrounds and extend these types of searches to low $X_{ee}$ masses.
We note that with the hadronic beam at NA62, the number of pion and kaon decays are comparable.
Even with a down-scaled trigger, NA62 may be well poised to perform such a search alongside other exotic channels like $K^+ \to e^+ \nu X_{ee}$.
A persistent $X(17)$ anomaly would also motivate a new set of $\pi^-$ capture experiments that would move the suggested anomaly to smaller angles ($\sim 16$ degrees) and be free from nuclear uncertainties~\cite{Chen:2019ivz}.
Finally, muon decays could also provide further insight.
Previous studies show that the Mu3e experiment at PSI can be sensitive to $Q_e^V \varepsilon$ couplings as low as $10^{-4}$ by searching for resonances in $\mu^+ \to e^+ \overline{\nu}_\mu \nu_e (X \to e^+e^-)$~\cite{Echenard:2014lma}.

\begin{acknowledgments}
MP is supported in part by U.S. Department of Energy (Grant No. desc0011842).
This research was supported in part by Perimeter Institute for Theoretical Physics. 
Research at Perimeter Institute is supported by the Government of Canada through the Department of Innovation, Science and Economic Development and by the Province of Ontario through the Ministry of Research, Innovation and Science.
\end{acknowledgments}

\appendix



\section{X17 in chiral pertubation theory}
\label{app:x17_chpt}

To calculate the rate for $\pi^+ \to \ell^+ \nu_\ell X$, we add the gauge boson $X_\mu$ as an external gauge field to the SU$(2)$ chiral perturbation theory (ChPT).
We include $X_\mu$ with both vector and axial-vector couplings to quarks and the electroweak bosons.
As usual, the gauge bosons are split into left-chiral and right-chiral gauge fields in the covariant derivative,
\begin{align}
    \mathcal{D}_\mu U = \partial_\mu U + i U \ell_\mu - i r_\mu U,
\end{align}
where $U = e^{i \Phi/F}$, with the usual SU$(2)$ representation for the Goldstone fields $\vec{\phi} = (\phi_1, \phi_2, \phi_3)$,
\begin{equation}
    \Phi = \vec{\phi} \cdot \vec{\tau}  =
    \sqrt{2} \left( \pi^+ \tau_+ + \pi^- \tau_- \right) + \pi^0 \tau_3,
\end{equation}
where $\vec{\tau} = (\tau_1, \tau_2, \tau_3)$ is the vector of Pauli matrices and $\tau_\pm = (\tau_1 \pm i\tau_2)/2$.
Here, $F = F_\pi \simeq 93$~MeV and we expand $U \simeq \mathbb{1} + i \Phi/F + \mathcal{O}(\Phi^2/F^2)$.
The left-chiral, $\ell_\mu \equiv v_\mu - a_\mu$, and right-chiral gauge field, $r_\mu \equiv v_\mu + a_\mu$, are given by
\begin{align}
    \ell_\mu &= \ell_\mu^X + \ell_\mu^W + \ell_\mu^Z, \quad 
    r_\mu = r_\mu^X + r_\mu^Z,
\end{align}
with 
\begin{align}
    \ell_\mu^X &= - e\varepsilon \left( \mathbb{1} Q_0^L + \tau_3 Q_3^L  \right) X_\mu,
    \\\nonumber
    r_\mu^X &= - e\varepsilon \left( \mathbb{1} Q_0^R + \tau_3 Q_3^R  \right) X_\mu,
    \\\nonumber
    \ell_\mu^Z &= \frac{g}{2c_W} \left(\frac{s_W^2}{6}\mathbb{1} - c_W^2\tau_3\right)  Z_\mu
    \\\nonumber
    r_\mu^Z &= \frac{g}{2c_W} \left(\frac{s_W^2}{6}\mathbb{1} + s_W^2\tau_3\right) Z_\mu
    \\\nonumber
    \ell_\mu^W &= -\frac{g}{\sqrt{2}} \left(V_{ud} W_\mu^+ \tau_+ + \text{h.c.}\right) 
\end{align}
where $V_{ud}$ is the first element of the quark mixing matrix.
We define the isoscalar and isovector quark charges under the $U(1)_X$,
\begin{align}
    Q^i_0 = \frac{Q^i_u + Q^i_d}{2}
    ,\quad 
    Q^i_3 = \frac{Q^i_u - Q^i_d}{2},
\end{align}
and $Q_i^L = Q_i^V - Q_i^A$ and $Q_i^R = Q_i^V + Q_i^A$.
 The photon, of course, also fits into this scheme
 and can be recovered from the general $X_\mu$ interactions by setting setting all axial-vector couplings to zero and $Q^V_0 \to 1/6$, $Q^V_3 \to 1/2$, $\varepsilon \to 1$.

From the $\mathcal{O}(p^2)$ ChPT Lagrangian, we collect the relevant interaction terms as
\begin{align}
    \mathscr{L}^{(2)} &= 
    \frac{F^2}{4} \langle D_\mu U (D^\mu U)^\dagger  \rangle  
    \\\nonumber &
    = \mathscr{L}_{VV} 
    + \mathscr{L}_{ \pi V} + \mathscr{L}_{ \pi\pi V}+ \mathscr{L}_{\pi VV} + \dots
\end{align}
The first term contains the masses for the gauge bosons generated by the quark condensate,
\begin{align}
&\mathscr{L}_{VV} = 
F^2 \bigg[ (e \varepsilon)^2 \left((Q_u^A)^2 + (Q_d^A)^2\right) X_\mu X^\mu
\\\nonumber
& - \frac{g}{c_W} e \varepsilon Q_0^A X_\mu Z^\mu + \frac{g^2}{8 c_W^2} Z_\mu Z^\mu + \frac{g^2|V_{ud}|^2}{4}W_\mu^+ {W^\mu}^- \bigg].
\end{align}
The mass mixing between $X_\mu$ and the $Z^\mu$ is negligible for our purposes.
Next, the derivative interactions of the gauge bosons with the pion fields,
\begin{align}
\mathscr{L}_{\pi V} &=  -\frac{F}{2} \bigg[ g V_{ud} W^+_\mu \partial^\mu \pi^- 
\\\nonumber
& \qquad \qquad
+ \partial^\mu \pi^0  \left(\frac{g}{2 c_W} Z_\mu - e \varepsilon Q_3^A X_\mu \right) \bigg] + \text{ h.c.},
\end{align}
and the vector current interaction,
\begin{align}
\mathscr{L}_{\pi\pi V} &= \pi^+ i\overset{\leftrightarrow}{\partial}_\mu\pi^- \left[ 2e\varepsilon Q_3^V X_\mu + \frac{g}{2 c_W} \cos{2\theta_W} Z_\mu\right].
\end{align}
The first term above is responsible for the bremsstrahlung of $X_\mu$ off the pion line in IB$_2$ in \cref{fig:diagrams}.
It is a purely vectorial interaction.
The contact terms responsible for the seagull diagrams in meson decay appear in
\begin{align}
\mathscr{L}_{ \pi VV} &=  ig e \varepsilon \bigg[ V_{ud}  Q_3^R (F + i\pi^0) \, X^\mu W_\mu^+ \pi^- 
\\\nonumber
&\qquad\qquad - \frac{Q_0^A F - Q_3^A \pi^0}{2 c_W} \, Z_\mu X^\mu \pi^0 \bigg] + \text{h.c.}
\end{align}
The first term above is responsible for IB$_3$ in \cref{fig:diagrams} and is proportional to the right-handed isovector quark charge.

\section{Pion decay amplitude}
\label{app:amplitude}

From the Lagrangian in the previous section, we compute the IB of $X_\mu$ in pion decays.
Following the diagrams in \cref{fig:diagrams}, the amplitude for 
\begin{equation}
\pi^+(k_1) \to e^+(k_2) \, \nu_e (k_3) \, X(k_4)
\end{equation}
is given by $\mathcal{M}^\mu_{\rm IB} = \mathcal{M}^\mu_1 + \mathcal{M}^\mu_2 + \mathcal{M}^\mu_3 + \mathcal{M}^\mu_4$ and can be rearranged into the form
\begin{align}
    \mathcal{M}_{\rm IB}  = \sqrt{2}& V_{ud}^*\,G_F\,F_\pi \, e \varepsilon \,
    \overline{u}(k_3) \Gamma^\mu \varepsilon_{\mu}^*(k_4) v(k_2),
\end{align}
where
\begin{align}
    \Gamma^\mu &= (Q_u^R - Q_d^R - Q_\nu^L + Q_e^L) \, \gamma^\mu P_L 
        \\\nonumber&
        + Q_\pi^V \frac{ 2 k_1^\mu}{k_{23}^2 - m_\pi^2} (m_\nu P_L - m_e P_R)
        \\\nonumber
        & + \frac{\slashed{k}_{24} \gamma^\mu}{k_{24}^2 - m_e^2}(m_\nu Q^L_e P_L - m_e Q_e^R P_R)
        \\\nonumber&
        - \frac{m_e \gamma^\mu}{k_{24}^2 - m_e^2}(m_\nu Q^R_e P_R - m_e Q_e^L P_L)
        \\\nonumber
        & - \frac{\gamma^\mu \slashed{k}_{34}}{k_{34}^2 - m_\nu^2}(m_\nu Q^R_\nu P_L - m_e Q_\nu^L P_R)
        \\\nonumber&
        -  \frac{m_\nu \gamma^\mu}{k_{34}^2 - m_\nu^2}(m_\nu Q^L_\nu P_L - m_e Q_\nu^R P_R).
\end{align}
In the analogous EM process $\pi^+ \to e^+ \nu_e \gamma$, the total amplitude is helicity suppressed due to an exact cancellation between terms in $\mathcal{M}_1$ and $\mathcal{M}_3$, a fact guaranteed by gauge invariance.
The amplitude $\mathcal{M}_2$ is always suppressed by $m_e$.
This picture is unchanged for the $X$ boson, provided the current to which $X$ couples is conserved.
In particular, the $m_e$-unsuppressed $X$ emission terms all cancel in the sum $\mathcal{M}_1 + \mathcal{M}_3 +\mathcal{M}_4$, provided $\Delta Q_X = Q_u^R - Q_d^R - (Q_\nu^L - Q_e^L) = 0$.

\paragraph{Structure dependent term}
The additional diagrams in \cref{fig:diagrams} come from the vector and axial-vector form factors of the pion, and can be directly related to $\pi^0 \to \gamma \gamma$ and the pion electromagnetic radius in the case of QED.
Most notably, the ratio of axial-vector to vector form factors, $\gamma = F_A(0)/F_V(0)$ was the subject of several theoretical and experimental efforts, eventually confirming the approach of chiral symmetry~\cite{Holstein:1986uj,Piilonen:1986bv,Bay:1986kf}.
In the case of a new vector or axial-vector boson, no such data-driven relation exists, and one has to resort to a microscopic model for the form factors. 
It is well-known that the vector and axial-vector SD terms are dominated by the exchange of vector and axial-vector mesons, in particular, in the framework of vector meson dominance (VMD), by the emission of $\rho$, $\omega$, and $\phi$, and their subsequent mixing with the photon.
Since $X(17)$ is not a strongly-interacting particle, its emission from the SD terms proceeds exclusively via its mixing with vector and axial-vector mesons.
Switching to the SU$(3)$ version of ChPT and following the hidden gauge symmetry approach to VMD~\cite{Bando:1984ej}, the relevant mixing terms are given by
\begin{align}
    \mathscr{L}_{XV} &= - M_V^2 \frac{e \varepsilon}{g} X_\mu\langle Q^V V^\mu \rangle 
    \\ \nonumber
    &= - 2 M_V^2 \frac{e \varepsilon}{g} X_\mu \left( Q_3^V \rho^\mu  + Q_0^V \omega^\mu + Q_s^V \phi^\mu\right)
\end{align}
where $g = M_V/2F$ is the hidden symmetry gauge coupling and $V^\mu$ the SU$(3)$ vector meson matrix.
The above allows the evaluation of the SD diagrams in \cref{fig:diagrams}, and since $M_V \gg m_X$, we see no particular enhancement to this rate with respect to the QED case for a $17$~MeV boson. 
A similar argument holds for the axial-vector case, where $X_\mu$ can mix with $a_1$, for instance.
Since the SD piece is already very small for the helicity-suppressed case, we can conclude that it can be safely neglected in our discussion.

\bibliographystyle{apsrev4-1}
\bibliography{main}{}

\end{document}